\documentclass[aps,prl,twocolumn]{revtex4}  \usepackage{graphicx}
\usepackage{amsmath}

\def\PsfigVersion{1.9}
\ifx\undefined\psfig\else \fi

\let\LaTeXAtSign=\@
\let\@=\relax
\edef\psfigRestoreAt{\catcode`\@=\number\catcode`@\relax}
\catcode`\@=11\relax
\newwrite\@unused
\def\ps@typeout#1{{\let\protect\string\immediate\write\@unused{#1}}}
\ps@typeout{psfig/tex \PsfigVersion}

\def\figurepath{./}

\def\@nnil{\@nil}
\def\@empty{}
\def\@psdonoop#1\@@#2#3{}
\def\@psdo#1:=#2\do#3{\edef\@psdotmp{#2}\ifx\@psdotmp\@empty \else
    \expandafter\@psdoloop#2,\@nil,\@nil\@@#1{#3}\fi}
\def\@psdoloop#1,#2,#3\@@#4#5{\def#4{#1}\ifx #4\@nnil \else
       #5\def#4{#2}\ifx #4\@nnil \else#5\@ipsdoloop #3\@@#4{#5}\fi\fi}
\def\@ipsdoloop#1,#2\@@#3#4{\def#3{#1}\ifx #3\@nnil 
       \let\@nextwhile=\@psdonoop \else
      #4\relax\let\@nextwhile=\@ipsdoloop\fi\@nextwhile#2\@@#3{#4}}
\def\@tpsdo#1:=#2\do#3{\xdef\@psdotmp{#2}\ifx\@psdotmp\@empty \else
    \@tpsdoloop#2\@nil\@nil\@@#1{#3}\fi}
\def\@tpsdoloop#1#2\@@#3#4{\def#3{#1}\ifx #3\@nnil 
       \let\@nextwhile=\@psdonoop \else
      #4\relax\let\@nextwhile=\@tpsdoloop\fi\@nextwhile#2\@@#3{#4}}
\ifx\undefined\fbox
\newdimen\fboxrule
\newdimen\fboxsep
\newdimen\ps@tempdima
\newbox\ps@tempboxa
\long\def\fbox#1{\leavevmode\setbox\ps@tempboxa\hbox{#1}\ps@tempdima\fboxrule
    \advance\ps@tempdima \fboxsep \advance\ps@tempdima \dp\ps@tempboxa
   \hbox{\lower \ps@tempdima\hbox
  {\vbox{\hrule height \fboxrule
          \hbox{\vrule width \fboxrule \hskip\fboxsep
          \vbox{\vskip\fboxsep \box\ps@tempboxa\vskip\fboxsep}\hskip 
                 \fboxsep\vrule width \fboxrule}
                 \hrule height \fboxrule}}}}
\fi
\newread\ps@stream
\newif\ifnot@eof       
\newif\if@noisy        
\newif\if@atend        
\newif\if@psfile       
{\catcode`\%=12\global\gdef\epsf@start{
\def\epsf@PS{PS}
\def\epsf@getbb#1{%
\openin\ps@stream=#1
\ifeof\ps@stream\ps@typeout{Error, File #1 not found}\else
   {\not@eoftrue \chardef\other=12
    \def\do##1{\catcode`##1=\other}\dospecials \catcode`\ =10
    \loop
       \if@psfile
	  \read\ps@stream to \epsf@fileline
       \else{
	  \obeyspaces
          \read\ps@stream to \epsf@tmp\global\let\epsf@fileline\epsf@tmp}
       \fi
       \ifeof\ps@stream\not@eoffalse\else
       \if@psfile\else
       \expandafter\epsf@test\epsf@fileline:. \\%
       \fi
          \expandafter\epsf@aux\epsf@fileline:. \\%
       \fi
   \ifnot@eof\repeat
   }\closein\ps@stream\fi}%
\long\def\epsf@test#1#2#3:#4\\{\def\epsf@testit{#1#2}
			\ifx\epsf@testit\epsf@start\else
\ps@typeout{Warning! File does not start with `\epsf@start'.  It may not be a PostScript file.}
			\fi
			\@psfiletrue} 
{\catcode`\%=12\global\let\epsf@percent=
\long\def\epsf@aux#1#2:#3\\{\ifx#1\epsf@percent
   \def\epsf@testit{#2}\ifx\epsf@testit\epsf@bblit
	\@atendfalse
        \epsf@atend #3 . \\%
	\if@atend	
	   \if@verbose{
		\ps@typeout{psfig: found `(atend)'; continuing search}
	   }\fi
        \else
        \epsf@grab #3 . . . \\%
        \not@eoffalse
        \global\no@bbfalse
        \fi
   \fi\fi}%
\def\epsf@grab #1 #2 #3 #4 #5\\{%
   \global\def\epsf@llx{#1}\ifx\epsf@llx\empty
      \epsf@grab #2 #3 #4 #5 .\\\else
   \global\def\epsf@lly{#2}%
   \global\def\epsf@urx{#3}\global\def\epsf@ury{#4}\fi}%
\def\epsf@atendlit{(atend)} 
\def\epsf@atend #1 #2 #3\\{%
   \def\epsf@tmp{#1}\ifx\epsf@tmp\empty
      \epsf@atend #2 #3 .\\\else
   \ifx\epsf@tmp\epsf@atendlit\@atendtrue\fi\fi}

\chardef\psletter = 11 
\chardef\other = 12

\newif \ifdebug 
\newif\ifc@mpute 
\c@mputetrue 

\let\then = \relax
\def\r@dian{pt }
\let\r@dians = \r@dian
\let\dimensionless@nit = \r@dian
\let\dimensionless@nits = \dimensionless@nit
\def\internal@nit{sp }
\let\internal@nits = \internal@nit
\newif\ifstillc@nverging
\def \Mess@ge #1{\ifdebug \then \message {#1} \fi}

{ 
	\catcode `\@ = \psletter
	\gdef \nodimen {\expandafter \n@dimen \the \dimen}
	\gdef \term #1 #2 #3%
	       {\edef \t@ {\the #1}
		\edef \t@@ {\expandafter \n@dimen \the #2\r@dian}%
		\t@rm {\t@} {\t@@} {#3}%
	       }
	\gdef \t@rm #1 #2 #3%
	       {{%
		\count 0 = 0
		\dimen 0 = 1 \dimensionless@nit
		\dimen 2 = #2\relax
		\Mess@ge {Calculating term #1 of \nodimen 2}%
		\loop
		\ifnum	\count 0 < #1
		\then	\advance \count 0 by 1
			\Mess@ge {Iteration \the \count 0 \space}%
			\Multiply \dimen 0 by {\dimen 2}%
			\Mess@ge {After multiplication, term = \nodimen 0}%
			\Divide \dimen 0 by {\count 0}%
			\Mess@ge {After division, term = \nodimen 0}%
		\repeat
		\Mess@ge {Final value for term #1 of 
				\nodimen 2 \space is \nodimen 0}%
		\xdef \Term {#3 = \nodimen 0 \r@dians}%
		\aftergroup \Term
	       }}
	\catcode `\p = \other
	\catcode `\t = \other
	\gdef \n@dimen #1pt{#1} 
}

\def \Divide #1by #2{\divide #1 by #2} 

\def \Multiply #1by #2
       {{
	\count 0 = #1\relax
	\count 2 = #2\relax
	\count 4 = 65536
	\Mess@ge {Before scaling, count 0 = \the \count 0 \space and
			count 2 = \the \count 2}%
	\ifnum	\count 0 > 32767 
	\then	\divide \count 0 by 4
		\divide \count 4 by 4
	\else	\ifnum	\count 0 < -32767
		\then	\divide \count 0 by 4
			\divide \count 4 by 4
		\else
		\fi
	\fi
	\ifnum	\count 2 > 32767 
	\then	\divide \count 2 by 4
		\divide \count 4 by 4
	\else	\ifnum	\count 2 < -32767
		\then	\divide \count 2 by 4
			\divide \count 4 by 4
		\else
		\fi
	\fi
	\multiply \count 0 by \count 2
	\divide \count 0 by \count 4
	\xdef \product {#1 = \the \count 0 \internal@nits}%
	\aftergroup \product
       }}

\def\r@duce{\ifdim\dimen0 > 90\r@dian \then   
		\multiply\dimen0 by -1
		\advance\dimen0 by 180\r@dian
		\r@duce
	    \else \ifdim\dimen0 < -90\r@dian \then  
		\advance\dimen0 by 360\r@dian
		\r@duce
		\fi
	    \fi}

\def\Sine#1%
       {{%
	\dimen 0 = #1 \r@dian
	\r@duce
	\ifdim\dimen0 = -90\r@dian \then
	   \dimen4 = -1\r@dian
	   \c@mputefalse
	\fi
	\ifdim\dimen0 = 90\r@dian \then
	   \dimen4 = 1\r@dian
	   \c@mputefalse
	\fi
	\ifdim\dimen0 = 0\r@dian \then
	   \dimen4 = 0\r@dian
	   \c@mputefalse
	\fi
	\ifc@mpute \then
		\divide\dimen0 by 180
		\dimen0=3.141592654\dimen0
		\dimen 2 = 3.1415926535897963\r@dian 
		\divide\dimen 2 by 2 
		\Mess@ge {Sin: calculating Sin of \nodimen 0}%
		\count 0 = 1 
		\dimen 2 = 1 \r@dian 
		\dimen 4 = 0 \r@dian 
		\loop
			\ifnum	\dimen 2 = 0 
			\then	\stillc@nvergingfalse 
			\else	\stillc@nvergingtrue
			\fi
			\ifstillc@nverging 
			\then	\term {\count 0} {\dimen 0} {\dimen 2}%
				\advance \count 0 by 2
				\count 2 = \count 0
				\divide \count 2 by 2
				\ifodd	\count 2 
				\then	\advance \dimen 4 by \dimen 2
				\else	\advance \dimen 4 by -\dimen 2
				\fi
		\repeat
	\fi		
			\xdef \sine {\nodimen 4}%
       }}

\def\Cosine#1{\ifx\sine\UnDefined\edef\Savesine{\relax}\else
		             \edef\Savesine{\sine}\fi
	{\dimen0=#1\r@dian\advance\dimen0 by 90\r@dian
	 \Sine{\nodimen 0}
	 \xdef\cosine{\sine}
	 \xdef\sine{\Savesine}}}	      

\def\psdraft{
	\def\@psdraft{0}
}
\def\psfull{
	\def\@psdraft{100}
}

\psfull

\newif\if@scalefirst
\def\psscalefirst{\@scalefirsttrue}
\def\psrotatefirst{\@scalefirstfalse}
\psrotatefirst

\newif\if@draftbox
\def\psnodraftbox{
	\@draftboxfalse
}
\def\psdraftbox{
	\@draftboxtrue
}
\@draftboxtrue

\newif\if@prologfile
\newif\if@postlogfile
\def\pssilent{
	\@noisyfalse
}
\def\psnoisy{
	\@noisytrue
}
\psnoisy
\newif\if@bbllx
\newif\if@bblly
\newif\if@bburx
\newif\if@bbury
\newif\if@height
\newif\if@width
\newif\if@rheight
\newif\if@rwidth
\newif\if@angle
\newif\if@clip
\newif\if@verbose
\newif\if@scale
\def\@p@@sclip#1{\@cliptrue}

\newif\if@decmpr

\def\@p@@sfigure#1{\def\@p@sfile{null}\def\@p@sbbfile{null}
	        \openin1=#1.bb
		\ifeof1\closein1
	        	\openin1=\figurepath#1.bb
			\ifeof1\closein1
			        \openin1=#1
				\ifeof1\closein1%
				       \openin1=\figurepath#1
					\ifeof1
					   \ps@typeout{Error, File #1 not found}
						\if@bbllx\if@bblly
				   		\if@bburx\if@bbury
			      				\def\@p@sfile{#1}%
			      				\def\@p@sbbfile{#1}%
							\@decmprfalse
				  	   	\fi\fi\fi\fi
					\else\closein1
				    		\def\@p@sfile{\figurepath#1}%
				    		\def\@p@sbbfile{\figurepath#1}%
						\@decmprfalse
	                       		\fi%
			 	\else\closein1%
					\def\@p@sfile{#1}
					\def\@p@sbbfile{#1}
					\@decmprfalse
			 	\fi
			\else
				\def\@p@sfile{\figurepath#1}
				\def\@p@sbbfile{\figurepath#1.bb}
				\@decmprtrue
			\fi
		\else
			\def\@p@sfile{#1}
			\def\@p@sbbfile{#1.bb}
			\@decmprtrue
		\fi}

\def\@p@@sfile#1{\@p@@sfigure{#1}}

\def\@p@@sbbllx#1{
		\@bbllxtrue
		\dimen100=#1
		\edef\@p@sbbllx{\number\dimen100}
}
\def\@p@@sbblly#1{
		\@bbllytrue
		\dimen100=#1
		\edef\@p@sbblly{\number\dimen100}
}
\def\@p@@sbburx#1{
		\@bburxtrue
		\dimen100=#1
		\edef\@p@sbburx{\number\dimen100}
}
\def\@p@@sbbury#1{
		\@bburytrue
		\dimen100=#1
		\edef\@p@sbbury{\number\dimen100}
}
\def\@p@@sheight#1{
		\@heighttrue
		\dimen100=#1
   		\edef\@p@sheight{\number\dimen100}
}
\def\@p@@swidth#1{
		\@widthtrue
		\dimen100=#1
		\edef\@p@swidth{\number\dimen100}
}
\def\@p@@srheight#1{
		\@rheighttrue
		\dimen100=#1
		\edef\@p@srheight{\number\dimen100}
}
\def\@p@@srwidth#1{
		\@rwidthtrue
		\dimen100=#1
		\edef\@p@srwidth{\number\dimen100}
}
\def\@p@@sangle#1{
		\@angletrue
		\edef\@p@sangle{#1} 
}
\def\@p@@srotate#1{\@p@@sangle{-#1}}
\def\@p@@sscale#1{
		\@scaletrue
		\edef\@p@sscale{#1}
}
\def\@p@@ssilent#1{ 
		\@verbosefalse
}
\def\@p@@sprolog#1{\@prologfiletrue\def\@prologfileval{#1}}
\def\@p@@spostlog#1{\@postlogfiletrue\def\@postlogfileval{#1}}
\def\@cs@name#1{\csname #1\endcsname}
\def\@setparms#1=#2,{\@cs@name{@p@@s#1}{#2}}
\def\ps@init@parms{
		\@bbllxfalse \@bbllyfalse
		\@bburxfalse \@bburyfalse
		\@heightfalse \@widthfalse
		\@rheightfalse \@rwidthfalse
		\@scalefalse
		\def\@p@sbbllx{}\def\@p@sbblly{}
		\def\@p@sbburx{}\def\@p@sbbury{}
		\def\@p@sheight{}\def\@p@swidth{}
		\def\@p@srheight{}\def\@p@srwidth{}
		\def\@p@sangle{0}
		\def\@p@sfile{} \def\@p@sbbfile{}
		\def\@p@scost{10}
		\def\@sc{}
		\@prologfilefalse
		\@postlogfilefalse
		\@clipfalse
		\if@noisy
			\@verbosetrue
		\else
			\@verbosefalse
		\fi
}
\def\parse@ps@parms#1{
	 	\@psdo\@psfiga:=#1\do
		   {\expandafter\@setparms\@psfiga,}}
\newif\ifno@bb
\def\bb@missing{
	\if@verbose{
		\ps@typeout{psfig: searching \@p@sbbfile \space  for bounding box}
	}\fi
	\no@bbtrue
	\epsf@getbb{\@p@sbbfile}
        \ifno@bb \else \bb@cull\epsf@llx\epsf@lly\epsf@urx\epsf@ury\fi
}	
\def\bb@cull#1#2#3#4{
	\dimen100=#1 bp\edef\@p@sbbllx{\number\dimen100}
	\dimen100=#2 bp\edef\@p@sbblly{\number\dimen100}
	\dimen100=#3 bp\edef\@p@sbburx{\number\dimen100}
	\dimen100=#4 bp\edef\@p@sbbury{\number\dimen100}
	\no@bbfalse
}
\newdimen\p@intvaluex
\newdimen\p@intvaluey
\def\rotate@#1#2{{\dimen0=#1 sp\dimen1=#2 sp
		  \global\p@intvaluex=\cosine\dimen0
		  \dimen3=\sine\dimen1
		  \global\advance\p@intvaluex by -\dimen3
		  \global\p@intvaluey=\sine\dimen0
		  \dimen3=\cosine\dimen1
		  \global\advance\p@intvaluey by \dimen3
		  }}
\def\compute@bb{
		\no@bbfalse
		\if@bbllx \else \no@bbtrue \fi
		\if@bblly \else \no@bbtrue \fi
		\if@bburx \else \no@bbtrue \fi
		\if@bbury \else \no@bbtrue \fi
		\ifno@bb \bb@missing \fi
		\ifno@bb \ps@typeout{FATAL ERROR: no bb supplied or found}
			\no-bb-error
		\fi
		\count203=\@p@sbburx
		\count204=\@p@sbbury
		\advance\count203 by -\@p@sbbllx
		\advance\count204 by -\@p@sbblly
		\edef\ps@bbw{\number\count203}
		\edef\ps@bbh{\number\count204}
		\if@angle 
			\Sine{\@p@sangle}\Cosine{\@p@sangle}
	        	{\dimen100=\maxdimen\xdef\r@p@sbbllx{\number\dimen100}
					    \xdef\r@p@sbblly{\number\dimen100}
			                    \xdef\r@p@sbburx{-\number\dimen100}
					    \xdef\r@p@sbbury{-\number\dimen100}}
                        \def\minmaxtest{
			   \ifnum\number\p@intvaluex<\r@p@sbbllx
			      \xdef\r@p@sbbllx{\number\p@intvaluex}\fi
			   \ifnum\number\p@intvaluex>\r@p@sbburx
			      \xdef\r@p@sbburx{\number\p@intvaluex}\fi
			   \ifnum\number\p@intvaluey<\r@p@sbblly
			      \xdef\r@p@sbblly{\number\p@intvaluey}\fi
			   \ifnum\number\p@intvaluey>\r@p@sbbury
			      \xdef\r@p@sbbury{\number\p@intvaluey}\fi
			   }
			\rotate@{\@p@sbbllx}{\@p@sbblly}
			\minmaxtest
			\rotate@{\@p@sbbllx}{\@p@sbbury}
			\minmaxtest
			\rotate@{\@p@sbburx}{\@p@sbblly}
			\minmaxtest
			\rotate@{\@p@sbburx}{\@p@sbbury}
			\minmaxtest
			\edef\@p@sbbllx{\r@p@sbbllx}\edef\@p@sbblly{\r@p@sbblly}
			\edef\@p@sbburx{\r@p@sbburx}\edef\@p@sbbury{\r@p@sbbury}
		\fi
		\count203=\@p@sbburx
		\count204=\@p@sbbury
		\advance\count203 by -\@p@sbbllx
		\advance\count204 by -\@p@sbblly
		\edef\@bbw{\number\count203}
		\edef\@bbh{\number\count204}
}
\def\in@hundreds#1#2#3{\count240=#2 \count241=#3
		     \count100=\count240	
		     \divide\count100 by \count241
		     \count101=\count100
		     \multiply\count101 by \count241
		     \advance\count240 by -\count101
		     \multiply\count240 by 10
		     \count101=\count240	
		     \divide\count101 by \count241
		     \count102=\count101
		     \multiply\count102 by \count241
		     \advance\count240 by -\count102
		     \multiply\count240 by 10
		     \count102=\count240	
		     \divide\count102 by \count241
		     \count200=#1\count205=0
		     \count201=\count200
			\multiply\count201 by \count100
		 	\advance\count205 by \count201
		     \count201=\count200
			\divide\count201 by 10
			\multiply\count201 by \count101
			\advance\count205 by \count201
		     \count201=\count200
			\divide\count201 by 100
			\multiply\count201 by \count102
			\advance\count205 by \count201
		     \edef\@result{\number\count205}
}
\def\ps@scaleinhundreds#1{
		\in@hundreds{#1}{\@p@sscale}{100}
		\edef#1{\@result}
}
\def\compute@wfromh{
		\in@hundreds{\@p@sheight}{\@bbw}{\@bbh}
		\edef\@p@swidth{\@result}
}
\def\compute@hfromw{
	        \in@hundreds{\@p@swidth}{\@bbh}{\@bbw}
		\edef\@p@sheight{\@result}
}
\def\compute@handw{
		\if@height 
			\if@width
			\else
				\compute@wfromh
			\fi
		\else 
			\if@width
				\compute@hfromw
			\else
				\edef\@p@sheight{\@bbh}
				\edef\@p@swidth{\@bbw}
			\fi
		\fi
}
\def\compute@resv{
		\if@rheight \else \edef\@p@srheight{\@p@sheight} \fi
		\if@rwidth \else \edef\@p@srwidth{\@p@swidth} \fi
}
\def\compute@sizes{
	\compute@bb
	\if@scalefirst\if@angle
	\if@width
	   \in@hundreds{\@p@swidth}{\@bbw}{\ps@bbw}
	   \edef\@p@swidth{\@result}
	\fi
	\if@height
	   \in@hundreds{\@p@sheight}{\@bbh}{\ps@bbh}
	   \edef\@p@sheight{\@result}
	\fi
	\fi\fi
	\compute@handw
	\compute@resv
	\if@scale
	   \if@verbose
	      \ps@typeout{(scaling by \@p@sscale)}%
	   \fi
	   \ps@scaleinhundreds{\@p@swidth}%
	   \ps@scaleinhundreds{\@p@sheight}%
	   \ps@scaleinhundreds{\@p@srwidth}%
	   \ps@scaleinhundreds{\@p@srheight}%
	\fi
}

\def\psfig#1{\vbox {
	\ps@init@parms
	\parse@ps@parms{#1}
	\compute@sizes
	\ifnum\@p@scost<\@psdraft{
		\special{ps::[begin] 	\@p@swidth \space \@p@sheight \space
				\@p@sbbllx \space \@p@sbblly \space
				\@p@sbburx \space \@p@sbbury \space
				startTexFig \space }
		\if@angle
			\special {ps:: \@p@sangle \space rotate \space} 
		\fi
		\if@clip{
			\if@verbose{
				\ps@typeout{(clip)}
			}\fi
			\special{ps:: doclip \space }
		}\fi
		\if@prologfile
		    \special{ps: plotfile \@prologfileval \space } \fi
		\if@decmpr{
			\if@verbose{
				\ps@typeout{psfig: including \@p@sfile.Z \space }
			}\fi
			\special{ps: plotfile "`zcat \@p@sfile.Z" \space }
		}\else{
			\if@verbose{
				\ps@typeout{psfig: including \@p@sfile \space }
			}\fi
			\special{ps: plotfile \@p@sfile \space }
		}\fi
		\if@postlogfile
		    \special{ps: plotfile \@postlogfileval \space } \fi
		\special{ps::[end] endTexFig \space }
		\vbox to \@p@srheight true sp{
			\hbox to \@p@srwidth true sp{
				\hss
			}
		\vss
		}
	}\else{
		\if@draftbox{		
			\hbox{\frame{\vbox to \@p@srheight true sp{
			\vss
			\hbox to \@p@srwidth true sp{ \hss \@p@sfile \hss }
			\vss
			}}}
		}\else{
			\vbox to \@p@srheight true sp{
			\vss
			\hbox to \@p@srwidth true sp{\hss}
			\vss
			}
		}\fi

	}\fi
}}
\psfigRestoreAt
\let\@=\LaTeXAtSign

\begin{document}

\title{Power and group velocity}   \author{S. Selenu}

\affiliation{Atomistic Simulation Centre, School of Mathematics and
 Physics\\ Queen's University Belfast, Belfast BT7 1NN, Northern
 Ireland, UK}

\begin{abstract}
Here we make use of the Hellmann-Feynman theorem, with the aim to calculate
macroscopic quantum velocities instead of forces, and we show how it is possible 
to derive the expression of the work per unit time per unit volume (power density) done by a static external
uniform electromagnetic field interacting with a  quantal body. 
\end{abstract}

\date{\today} \maketitle

\section{Introduction}
\label{Geometric-phase}
\noindent The problem of the $\it{dielectric}$ and $\it{magnetic}$ response of 
matter, in condensed matter physics, have been intensively studied in the last
decades\cite{Nenciu}-\cite{Martin2} (and references therein).

Here we show that making use of a reformulation of the Hellmann-Feynman
theory, it is possible to derive several useful physical properties of a crystalline material in a direct and simple way.
As a matter of example we shall show how to derive the expression of the group velocity of an electronic system interacting 
with an external uniform magnetostatic field, and how to obtain the expression of the electronic 
power density.
 
Let us start our discussion from a fundamental level by briefly reviewing 
the definition of
$\it{stationary~state}$ or equivalently $\it{steady~state}$ of a quantal
body (QB), as better stated in \cite{Heisenberg,Schiff,Feynman}. 

Let us consider the time-dependent Schr$\ddot{o}$dinger equation

\begin{equation}
\label{steady}
\begin{split}
i\hbar\frac{\partial \Psi}{\partial t} &= -\frac{\hbar^{2}}{2m} \nabla_{{\bf{r}}}^{2} \Psi + V
({\bf{r}},t) \Psi \\&= (T+V) \Psi = H_{0} \Psi\\ 
\end{split}
\end{equation}

where $\Psi$ is the state of the system, $T$ the kinetic energy operator and
$V$ the potential energy.

A considerable simplification of equation
($\ref{steady}$) is obtained if the potential energy is time independent, i.e $V({\bf{r}},t)=V({\bf{r}})$.
In this case it is possible to express the general solution of
eq.(\ref{steady}) as a sum of products of functions of $\bf{r}$ and $t$ separately. 
Without entering into  details, the time dependence of the wave function can be expressed  as 

\begin{equation} 
\label{steadysol}
\Psi({\bf{r}},t)=u({\bf{r}}) e^{-i\frac{Et}{\hbar}}
\end{equation}

\noindent being $u({\bf{r}})$ the eigen function of the following differential equation \cite{Schiff}

\begin{equation} 
\label{steadysol1}
 -\frac{\hbar^{2}}{2m} \nabla_{{\bf{r}}}^{2} \Psi + V ({\bf{r}}) u({\bf{r}})=E u({\bf{r}}) 
\end{equation}

\noindent and $E$ is the corresponding eigen value. 

Eq.(\ref{steadysol1}) formally defines a stationary state (see also \cite{Heisenberg,Schiff,Feynman}) for a
system where the Hamiltonian assumes the form $H=T+V$. A more general
definition of  a steady state is 

\begin{equation}
\label{steady-states}
H\Psi=E\Psi
\end{equation}

where $H$ is the Hamiltonian of a quantal body, and $H$ is either
self-adjoint or hermitian (see \cite{Heisenberg,Schiff,Feynman}).
In the following we shall frequently make use of the concept of steady states. 

\section{Hellmann-Feynman Theorem}
\label{HF-theory}

In 1939 R.P. Feynman\cite{Feynman} developed formulas to calculate forces in a
quantum system, that nowadays take the name of Hellmann-Feynman theorem\cite{Feynman}. 

The theorem states that for a given configuration of nuclei, it is possible to calculate directly the
force required to hold them; 
where it is understood that the nuclei are to be held fixed in position as point charges. 
Also, given any arbitrary number of parameters $\lambda$
that specifies nuclear positions the expression of the force (in a
generalized sense\cite{Feynman}) is

\begin{equation}
\label{force1}
f_{\lambda}=-<\nabla_{\lambda}H>=-\int dv \Psi^{*}
\nabla_{\lambda} H \Psi  
\end{equation}
  
where $H$ is the (self-adjoint) Hamiltonian of the system, and $\Psi$ are
differentiable wave functions. 

Let us here re-state the theorem in order to avoid its repetition in the
text (see also \cite{Feynman}).

The wave function must be normalized at every $\lambda$, i.e. 

\begin{equation}
\label{normal}
(i) \int dv |\Psi|^{2} =1
\end{equation}

Steady state condition requires 

\begin{equation}
\label{steady-s}
(ii) H \Psi = E \Psi
\end{equation}

where $E$, $H$ and $\Psi$ have to be thought depending on $\lambda$. The
requirement that $H$ is a self-adjoint operator means 

\begin{equation}
\label{self-a}
(iii) \int dv  \Psi^{*} H \nabla_{\lambda} \Psi = \int dv
\nabla_{\lambda}\Psi H \Psi^{*}
\end{equation}

and it is also valid the following relation 

\begin{equation}
\label{comm1}
(iv) \frac{1}{i\hbar} [i\hbar \nabla_{\lambda},H] \Psi =\nabla_{\lambda} H \Psi
\end{equation}

The expectation value, is

\begin{equation}
\label{E-av}
E = \int dv \Psi^{*} H \Psi 
\end{equation}

that implies

\begin{equation}
\label{deriv} 
\nabla_{\lambda} E = \int dv \Psi^{*} \nabla_{\lambda}H \Psi + \int dv
\nabla_{\lambda} \Psi^{*} H \Psi + \int dv \Psi^{*} H \nabla_{\lambda} \Psi
\end{equation}
 
From (i),(ii),(iii), and (iv) we obtain

\begin{equation}
\label{deriv2} 
\nabla_{\lambda} E = \int dv \Psi^{*} \nabla_{\lambda}H \Psi =\int dv
\Psi^{*}\frac{1}{i\hbar} [i\hbar \nabla_{\lambda},H]\Psi  
\end{equation}

We shall show in the next section, how to make use of the theorem 
(\ref{deriv2}) from another perspective.

\section{Group velocity}
\label{finite-sample}

In this section it is shown how to calculate the group velocity of 
a quantal body.

We assume that the differentiable state $\Psi$ of the system,
varies parametrically with respect to a real vectorial field $\beta$.

We represent  $\Psi$ by complex differentiable functions, normalized in
a suitable region $D$ of space.

Also, we assume that the Hamiltonian $\hat{H}$ of the quantal body is
$\it{independent}$ of $\beta$ and
self-adjoint.

Let us define
a generalized group velocity vector as:

\begin{equation}
\label{vel-gruppo}
\bar{V_{\beta}}=\hbar^{-1}\nabla_{\beta}E(\beta)
\end{equation}

We consider here

\begin{equation}
\label{GenBloch}
\Psi_{\beta}=e^{i\bar{f}(\beta) \cdot {\bf{r}}} u_{\beta}
\end{equation}

where $\bar{f}(\beta)$ is a real vectorial field function of $\beta$.

$E$ is defined as follows

\begin{equation}
\label{Eb}
E(\beta)=<\Psi_{\beta}|\hat{H}|\Psi_{\beta}>=<u_{\beta}|\tilde{H}_{\beta}|u_{\beta}>=\int u_{\beta}^{*}
\tilde{H_{\beta}} u_{\beta} dv
\end{equation}

\noindent where $dv$ is a volume element of space, and $\tilde{H}_{\beta}$ is obtained as

\begin{equation}
\label{Hbeta}
\tilde{H}_{\beta}=e^{-i\bar{f}(\beta)\cdot {\bf{r}}}\hat{H}e^{i\bar{f}(\beta)\cdot {\bf{r}}}
\end{equation}

$\tilde{H}_{\beta}$ represents the Hamiltonian of the
system\cite{Dirac} in the Hilbert space spanned by $u_{\beta}$, and it is  parametrically dependent
on $\beta$.

Inserting eq.(\ref{Eb}) in eq.(\ref{vel-gruppo}) we obtain,

\begin{equation}
\label{Vk}
\hbar \bar{V_{\beta}}  =\nabla_{\beta}<\Psi_{\beta}|\hat{H}|\Psi_{\beta}> \\
 = \nabla_{\beta}<u_{\beta}|\tilde{H_{\beta}}|u_{\beta}>
\end{equation}

We  may consider $\bar{V_{\beta}}$ to be the averaged vector

\begin{equation}
\label{Vk0}
\bar{V_{\beta}}=\frac{1}{\hbar}<u_{\beta}|\nabla_{\beta} \tilde{H_{\beta}}|u_{\beta}>
\end{equation}

To prove that, under steady-state conditions, both  definitions
eq.(\ref{Vk0}) and eq(\ref{vel-gruppo}), of group velocity become
exactly equivalent, it is sufficient to make use of eq.(\ref{deriv2}),
where we consider here $\lambda$ being substituted by a Cartesian component of $\beta$.   
In the next section, as a matter of example, we show implications of
eq.(\ref{Vk0}) in the theory of infinite periodic systems.

\section{Infinite periodic system}
\label{Cristallo}
For an application of eq.(\ref{Vk0}), to an infinite periodic system, 
it is sufficient to make use of the $\it{k-q}$ representation,  introduced by
Zak\cite{Zak}. 

Here $\bf{k}$ is the crystal momentum (\cite{Ashcroft}) and varies in the
Brillouin zone of the reciprocal space while the quasi-coordinate $\bf{q}$ varies in
the unit cell.  

The $k-q$ representation of Schr$\ddot{o}$dinger's
equation, in the case of no external fields present in the system, is: 

\begin{equation}
\label{vel-kq}
[\frac{1}{2m}(-i\hbar \nabla_{{\bf{q}}})^{2} + V({\bf{q}})]
C_{{\bf{k}}}({\bf{q}}) = \epsilon C_{{\bf{k}}}({\bf{q}}) 
\end{equation} 

where

\begin{equation}
\label{H0}
H_{0} \equiv \frac{1}{2m}(-i\hbar \nabla_{{\bf{q}}})^{2} + V({\bf{q}})
\end{equation} 

\noindent as also stated in eq.(14) of reference \cite{Zak}, setting
explicitly to zero the external fields. 

Performing  the  following phase transformation    

\begin{equation}
\label{phase-transf}
C_{{\bf{k}}}({\bf{q}})=e^{i{\bf{k}} \cdot {\bf{q}}} u_{{\bf{k}}}({\bf{q}})
\end{equation}

the equation for $u_{\bf{k}}({\bf{q}})$ becomes 

\begin{equation}
\label{vel-kq1}
[\frac{1}{2m}(-i \hbar \nabla_{{\bf{q}}}+{\hbar \bf{k}})^{2} + V({\bf{q}})]
u_{{\bf{k}}}({\bf{q}}) = \epsilon u_{{\bf{k}}}({\bf{q}}) 
\end{equation} 

where 

\begin{equation}
\label{vel-kq2}
H_{0,{\bf{k}}} \equiv \frac{1}{2m}(-i \hbar \nabla_{{\bf{q}}}+{\hbar \bf{k}})^{2} + V({\bf{q}})
\end{equation} 

If we now consider a static external electromagnetic field $(E_{0},B^{0})$ interacting with the QB, 
Schr$\ddot{o}$dinger's equation in the $k-q$ representation becomes 

\begin{equation}
\label{vel-kq6}
\begin{split}
H_{{\bf{k}}}=&[\frac{1}{2m}(-i\hbar \nabla_{{\bf{q}}}+{\hbar \bf{k}}+\frac{e}{c}
  B^{0} \times i\nabla_{{\bf{k}}})^{2} \\ &+ V({\bf{q}}) + eE_{0}\cdot i\nabla_{{\bf{k}}}]
u_{{\bf{k}}}({\bf{q}})\\ &= \epsilon u_{{\bf{k}}}({\bf{q}})\\ 
\end{split}
\end{equation} 

differently from the result shown by Zak, because of our choice of the vector
potential. 

In fact, in the $\bf{r}$ representation, we choose the vector potential  as follows:

\begin{equation}
\label{vec-pot}
{\bf{A}}({\bf{r}})= B^{0} \times {\bf{r}}
\end{equation}

\noindent differently from the usual one used for the symmetric gauge
\cite{Landau-quantumM}. 

It is clear that we can apply straightforwardly eq.(\ref{Vk0}),
once we think of the region $D$ as being the super-cell,
and we substitute $\bar{f}({\beta})=\beta={\bf{k}}$, and ${\bf{r}}$ with
${\bf{q}}$.

We can also evaluate eq.(\ref{vel-kq6}), at low energies, expanding brackets
$(...)^{2}$ of eq.(\ref{vel-kq6}) in powers of $\frac{1}{c}$.

Taking only those terms proportional to $\frac{1}{c}$, we obtain

\begin{equation}
\label{vel-kq7}
\begin{split}
&[\frac{1}{2m}(-i\hbar \nabla_{{\bf{q}}}+{\hbar \bf{k}})^{2}+ \frac{e{\bf {v}}}{c}\cdot (B^{0} \times
    i\nabla_{{\bf{k}}}) \\ & + V({\bf{q}}) 
+ eE_{0}\cdot i\nabla_{{\bf{k}}}]
u_{{\bf{k}}}({\bf{q}}) \\&= \epsilon u_{{\bf{k}}}({\bf{q}})\\ 
\end{split}
\end{equation}

where ${\bf{v}}=\frac{1}{m}[-i \hbar \nabla_{{\bf{q}}}+{\hbar{\bf{k}}}]$,
and the Hamiltonian of the problem is 

\begin{equation}
\label{Hamiltoniana}
\begin{split}
\tilde{H_{{\bf{k}}}} & \equiv \frac{1}{2m}(-i\hbar \nabla_{{\bf{q}}}+{\hbar \bf{k}})^{2}+  \frac{e{\bf {v}}}{c} \cdot (B^{0} \times
    i\nabla_{{\bf{k}}}) \\ & + V({\bf{q}}) 
  + eE_{0}\cdot
    i\nabla_{{\bf{k}}}=H_{0,{\bf{k}}} + \hat{\omega} \\
\end{split}
\end{equation} 

The operator $\hat{\omega}$ is instead defined as

\begin{equation}
\label{omega}
\hat{\omega} =  eE_{0}\cdot i\nabla_{{\bf{k}}}  + \frac{e{\bf {v}}}{c} \cdot (B^{0} \times
    i\nabla_{{\bf{k}}}) 
\end{equation}

Boundary conditions on the Bloch-like functions are the following 

\begin{equation}
\label{BC}
\begin{split}
C_{{\bf{k}}+{\bf{G}_{m}}}({\bf{q}}) = C_{{\bf{k}}}({\bf{q}})\\
and\\
C_{{\bf{k}}}({\bf{q}+{\bf{R}_{m}}})=e^{i {\bf{k}} \cdot {{\bf{R}_{m}}}} C_{{\bf{k}}}({\bf{q}})\\
\end{split}
\end{equation}

being  ${\bf{G}_{m}}$ any reciprocal lattice vector and ${\bf{R}_{m}}$ any
lattice vector\cite{Ashcroft,Kittel,Martin-book}. 

Accordingly to (\ref{BC}), boundary conditions on functions $u_{\bf{k}}({\bf{q}})$ are 

\begin{equation}
\label{BC1}
\begin{split}
u_{{\bf{k}}}({\bf{q}}+{\bf{R}_{m}}) = u_{{\bf{k}}}({\bf{q}})\\
and\\
u_{{\bf{k}}+{\bf{G}_{m}}}({\bf{q}})=e^{-i {\bf{G}_{m}} \cdot {{\bf{q}}}} u_{{\bf{k}}}({\bf{q}})\\
\end{split}
\end{equation}

for every ${\bf{G}_{m}}$ and ${\bf{R}_{m}}$.

A particular solution then reads

\begin{equation}
\label{Gen-sol}
C_{\bf{k}}({\bf{q}}) = e^{i{\bf{k}} \cdot {\bf{q}}} u_{\bf{k}}({\bf{q}}) 
\end{equation}

Making use of eq.(\ref{Vk0}) and eq.(\ref{vel-kq6}), the expression of the group velocity operator now becomes 
 
\begin{equation}
\label{velocityG1B0}
\begin{split}
{\bf{V}}^{(B^{0})}&=\frac{1}{\hbar}\nabla_{\bf{k}} H_{{\bf{k}}}\\
&={\bf{v}} + \frac{e}{mc} B^{0} \times i\nabla_{\bf{k}} \\
\end{split}
\end{equation}

\section{Power} 

Here we show another possible application of eq.(\ref{velocityG1B0}).

By eq.(\ref{velocityG1B0}), we can calculate the group velocity of the system (in the case of
$B^{0}=0$) as follows 

\begin{equation}
\label{velocityG1}
\begin{split}
\tilde{V}_{n\bf{k}}&=\frac{1}{\hbar}<u_{n\bf{k}}|\nabla_{\bf{k}}
H_{\bf{k}}|u_{n\bf{k}}>\\
&= <u_{n\bf{k}}|\frac{1}{i\hbar}[i\nabla_{\bf{k}},H_{\bf{k}}]|u_{n\bf{k}}>\\
\end{split}
\end{equation}

defining the density current vector as\cite{Ashcroft} 
 
\begin{equation}
\label{current}
j_{n\bf{k}}=e\tilde{V}_{n\bf{k}}
\end{equation}  

We can then define the work, per unit time and per unit volume, 
made by the external electric field $E_{0}$ on the QB as

\begin{equation}
\label{power}
\begin{split}
W&=E_{0}\cdot J=E_{0}\cdot \sum_{n} \int_{\Omega_{\bf{k}_{F}}} \frac{d\bf{k}}{(2 \pi)^{3}} j_{n\bf{k}}\\
&=eE_{0}\cdot \sum_{n} \int_{\Omega_{\bf{k}_{F}}} \frac{d\bf{k}}{(2 \pi)^{3}} \tilde{V}_{n\bf{k}} \\
&=eE_{0}\cdot \sum_{n} \int_{\Omega_{\bf{k}_{F}}} \frac{d\bf{k}}{(2 \pi)^{3}}\frac{1}{\hbar}\nabla_{\bf{k}}E_{n\bf{k}} \\
\end{split}
\end{equation}

where $\Omega_{\bf{k}_{F}}$ is the volume of ${\bf{k}}$ points such that their
associated energies are smaller then the Fermi energy. 
A generalization of eq.(\ref{power}) to the case $B^{0} \neq 0$ will be given
in the next section.

\section{Power in a uniform static electromagnetic field}
\label{potenza}

By eq.(\ref{velocityG1B0}), noting that the group velocity is not explicitly dependent on a uniform
electrostatic field $E_{0}$ even being eventually present in the system,
and 
bearing in mind eq.(\ref{current}) and eq.(\ref{power}), we can express the work per unit
time and per unit volume done by the electromagnetic field on the body as

\begin{equation}
\label{power1}
\begin{split}
W&=E_{0} \cdot J  =  E^{0} \cdot \int_{V} dV  {\bf{j}}({\bf{k}}) \\
&=e E^{0} \cdot \bar{{\bf{v}}} +  \frac{4 \pi e^{2}}{mc^{2}}
{\bf{S}} \cdot \bar{{\bf{d}}} \\
\end{split}
\end{equation}

being $\bf{S}=\frac{c}{4 \pi} (E_{0}\times B^{0})$ the Poynting
vector\cite{Jackson} associated to the electromagnetic field, and

\begin{equation}
\label{Pmn0}
{\bf{d}}=\sum_{n} <u_{n,{\bf{k}}}| i\nabla_{{\bf{k}}}|u_{n,{\bf{k}}}>
\end{equation}

\begin{equation}
\label{Pmn1}
{\bar{\bf{P}}}=e\bar{\bf{d}}=e\int_{V} \frac{dk}{(2 \pi)^{3}}{\bf{d}}
\end{equation}

where $e$ is the electronic charge, and ${\bar{\bf{P}}}$ is the macroscopic polarization\cite{Resta2}.
Eq.(\ref{Pmn0}) and eq.(\ref{Pmn1}) can be easily related to the dipole moment per unit volume of the system\cite{Resta2}, 
that is expressed as a sum of Berry's connections\cite{Resta2}.
A review of the physical and mathematical meaning of eq.(\ref{Pmn1}) can be found in \cite{KS-V,Resta2,Martin-book}.

\section{Conclusions}
\label{conclusioni}
 
Within our model, concepts of group velocity and power density are directly related. 
Expression (\ref{power1}) may be useful for calculations of the measured
power, adsorbed or released \cite{Pauli,Fermi-thermodynamics,Nye} by a
quantal body in a transformation connecting two of its steady states; it may also be useful for calculations of Joule's effect,
as the latter is directly related to the concept of power, and a
quantification of the $\it{heat}$ acquired or
released by a QB, along a phase transformation between two of their steady states. 
Eq.(\ref{power}), can be helpful for a more accurate theoretical description
of materials that show an hysteresis loop\cite{Lines-Glass},i.e during a phase
transformation that changes their electrostatic potential in analogy with\cite{Resta2}. 
As a matter of example, we may think to apply our results to the study of dielectric properties of materials appertaining to the
phenomenological $\it{polar}$ class \cite{Lines-Glass}). 

\bibliography{refs}

\end{document}